\documentclass[aps,prl,reprint,superscriptaddress,showpacs]{revtex4-1}

\usepackage{graphicx,amsmath,amssymb,ulem}
\usepackage{verbatim}
\usepackage{ragged2e}
\usepackage{color}

\begin{document}

\title{Orbital angular momentum modes of high-gain parametric down-conversion}

\author{Lina Beltran}
\affiliation{Max-Planck Institute for the Science of Light, \\  Staudtstrasse 1, Erlangen  D-91058, Germany}
\affiliation{University of Erlangen-N\"urnberg, Staudtstrasse 7/B2, 91058 Erlangen, Germany}
\author{Gaetano Frascella}
\affiliation{Max-Planck Institute for the Science of Light, \\  Staudtstrasse 1, Erlangen  D-91058, Germany}
\author{Angela M.~Perez}
\affiliation{Max-Planck Institute for the Science of Light, \\  Staudtstrasse 1, Erlangen  D-91058, Germany}
\affiliation{University of Erlangen-N\"urnberg, Staudtstrasse 7/B2, 91058 Erlangen, Germany}
\author{Robert Fickler}
\affiliation{Department of Physics,
University of Ottawa,\\ 25 Templeton Street, Ottawa, Ontario K1N 6N5, Canada}
\author{Polina R.~Sharapova}
\affiliation{Department of Physics, University of Paderborn, Warburger Stra\ss{}e 100, Paderborn D-33098, Germany}
\author{Mathieu Manceau}
\affiliation{Max-Planck Institute for the Science of Light, \\  Guenther-Scharowsky-Str. 1 / Bau 24, Erlangen  D-91058, Germany}
\author{Olga V.~Tikhonova}
\affiliation{Physics Department, Moscow State University, Leninskiye Gory 1-2, Moscow 119991, Russia}
\author{Robert W.~Boyd}
\affiliation{Department of Physics,
University of Ottawa,\\ 25 Templeton Street, Ottawa, Ontario K1N 6N5, Canada}
\author{Gerd Leuchs}
\affiliation{Max-Planck Institute for the Science of Light, \\  Guenther-Scharowsky-Str. 1 / Bau 24, Erlangen  D-91058, Germany}
\affiliation{University of Erlangen-N\"urnberg, Staudtstrasse 7/B2, 91058 Erlangen, Germany}
\author{Maria V.~Chekhova}
\affiliation{Max-Planck Institute for the Science of Light, \\  Guenther-Scharowsky-Str. 1 / Bau 24, Erlangen  D-91058, Germany}
\affiliation{University of Erlangen-N\"urnberg, Staudtstrasse 7/B2, 91058 Erlangen, Germany}
\affiliation{Physics Department, Moscow State University, Leninskiye Gory 1-2, Moscow 119991, Russia}

\begin{abstract}
Light beams with orbital angular momentum (OAM) are convenient carriers of quantum information. They can be also used for imparting rotational motion to particles and provide high resolution in imaging. Due to the conservation of OAM in parametric down-conversion (PDC), signal and idler photons generated at low gain have perfectly anti-correlated OAM values. It is interesting to study the OAM properties of high-gain PDC, where the same OAM modes can be populated with large, but correlated, numbers of photons. Here we investigate the OAM spectrum of high-gain PDC and show that the OAM mode content can be controlled by varying the pump power and the configuration of the source. In our experiment, we use a source consisting of two nonlinear crystals separated by an air gap. We discuss the OAM properties of PDC radiation emitted by this source and suggest possible modifications.
\end{abstract}

\maketitle

\section{Introduction}
Orbital angular momentum (OAM) of light, apart of being interesting for fundamental studies~\cite{Fickler2012}, presents a useful tool for imaging~\cite{Fuerhapter2005}, a convenient basis for information encoding~\cite{Malik2014}, and can be also used for light-matter interaction, in particular for imparting rotation to particles~\cite{Padgett2011}. Physically the OAM is a twist of the wavefront of a beam, or the dependence of the phase of light on the azimuthal angle~\cite{Allen1992}. Depending on how many times and in what direction the phase is changed over a $2\pi$ range, the OAM takes integer values of $0,\pm1,\pm2,$ etc. It is noteworthy that OAM is conserved in nonlinear optical processes such as harmonic generation~\cite{Dholakia1996} and parametric down-conversion (PDC)~\cite{Mair2001}. In the case of PDC, OAM conservation leads to anti-correlation between the OAM values of signal and idler photons. For instance, if PDC is obtained from a pump beam without OAM, the signal and idler photons have always opposite OAM values, which, together with the broad distribution of OAM for each photon taken separately, leads to the entanglement of photon pairs in OAM. This has been observed in experiments with weakly pumped PDC~\cite{Mair2001,Krenn2014}.

Strongly pumped PDC~\cite{Jedrkiewicz2004} as well as four-wave mixing (FWM)~\cite{Boyer2008} and modulational instability (MI)~\cite{Finger2015} lead to the generation of bright beams containing many photon pairs, - bright squeezed vacuum (BSV), which is nonclassical light with macroscopic photon numbers~\cite{Chekhova2015}. Its strong photon-number correlations~\cite{Jedrkiewicz2004,Bondani2007,Brida2009,Agafonov2010,Finger2015} enable observation of macroscopic polarization entanglement~\cite{Iskhakov2012} and can even violate Bell's inequalities~\cite{Rosolek2015}. Twin-beam BSV manifests macroscopic photon-number entanglement~\cite{Stobinska2012}, which can be used for encoding quantum information into photon numbers~\cite{Chekhova2015}. Because of the high brightness, BSV is expected to interact efficiently with material objects, and its efficiency for nonlinear optics has been already demonstrated~\cite{Jedrkiewicz2011}.

An important feature of BSV produced through PDC or FWM/MI is that it is highly multimode, both temporally and spatially~\cite{Jedrkiewicz2004,Boyer2008,Boyer2008Science,Agafonov2010}. In particular, the spatial mode spectrum contains modes with OAM~\cite{Boyer2008,Sharapova2015}, correlated pairwise, so that the mode with the OAM number $l$ has photon-number correlations with the $-l$ mode provided that the pump beam has no OAM. This feature is very useful for information encoding, since the high information capacity of twin beams due to the photon-number entanglement can be combined with the information capacity provided by OAM. Moreover, the axial resolution of OAM beams~\cite{Fickler2012} can be further increased by taking advantage of the statistical properties of BSV. For instance, one can prepare an OAM beam in a state with sub-Poissonian statistics by imposing a condition on its twin beam and applying the feedforward technique~\cite{Iskhakov2016}. Using the strong interaction of bright twin beams with matter, one can pass the OAM correlations to mechanical objects. Finally, the photon-number correlation in OAM modes will make twin-beam BSV very sensitive to mechanical rotation, similar to how spatial correlations make twin beams sensitive to spatial displacement~\cite{Treps2002}.

\begin{figure}[htbp]
\centering\includegraphics[width=8cm]{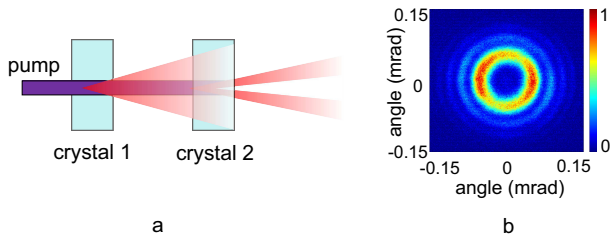}
\caption{The two-crystal scheme in the case of deamplification of collinear emission (a) and a typical single-shot far-field intensity distribution (b).}
\label{setup}
\end{figure}

In this work we investigate the possibility of controlling the OAM content of BSV. The scheme we use is based on PDC in two crystals placed one after another into a common pump beam (Fig.~\ref{setup}a). For a certain distance between the crystals (about $18$ mm), due to the small chromatic dispersion in the air gap between them, the emission from the first crystal in the collinear direction is deamplified in the second crystal~\cite{Perez2014}. At the same time, the radiation at nonzero angles can be amplified. As a result, the total emission pattern consists of concentric `donuts' (Fig.~\ref{setup}b). We analyze the OAM mode content of this radiation theoretically and measure it in experiment. In particular, we reconstruct the shapes and weights of the spatial eigenmodes through the statistical processing of single-shot spectra. For the description of spatial modes, we use the so-called Schmidt basis, in which the intermodal photon-number correlations are most simplified. The spatial Schmidt modes are two-dimensional, and we discuss separately their radial and azimuthal parts. The latter, being responsible for the OAM, are also called OAM modes and are the main subject of this work. Moreover, we vary the OAM content by changing the distance between the crystals as well as the pump power. Finally, we propose several ways to expand the tunability of this source with respect to the OAM spectrum.

\section{Methods}
\subsection{Experimental setup and methods}
\begin{figure}[htbp]
\centering\includegraphics[width=8.5cm]{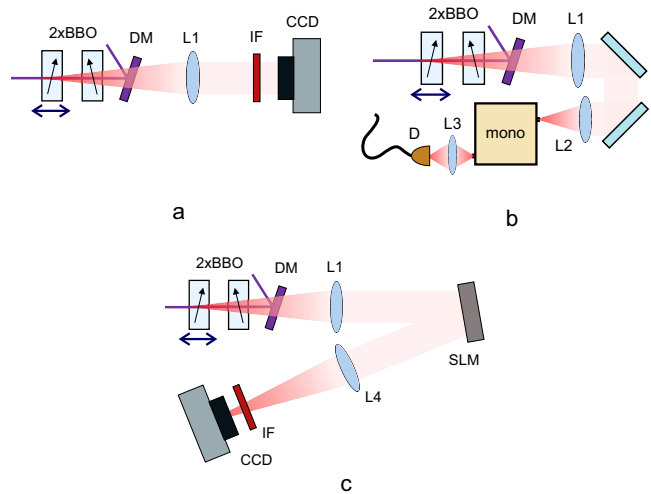}
\caption{Different versions of the experimental setup: measuring the far-field intensity distributions (a); measuring the total number of spatial modes (b) and measuring the OAM mode content by phase flattening (c). DM, dichroic mirror; D, analog detector; L1-L4, lenses; IF, interference filter; mono, monochromator, SLM, spatial light modulator.}
\label{experimental}
\end{figure}

The experimental setup is shown in Fig.~\ref{experimental}, with its three different versions depicted in panels a-c. The pump (third harmonic radiation of Nd:YAG laser, with the wavelength 354.67 nm, pulse duration 18 ps, and pulse repetition rate 1 kHz), is focused into a pair of type-I BBO crystals of length $L_c=2$ mm, placed at a distance $L$, which is variable from $7$ to $27$ mm. The pump beam waist has a full width at half maximum (FWHM) of $170\,\mu$m, which corresponds to the Rayleigh length $18$ cm. In this case, the pump beam diameter can be considered as constant between the two crystals. It is also worth noting that with this relatively soft  pump focusing, the spatial walk-off of the pump does not play a significant role, which leads to the averaged far-field intensity being almost independent on the azimuthal angle. Note that instantaneous (single-shot) intensity distributions still show variation with the azimuthal angle, but this variation is always centrally symmetric (see Fig.~\ref{setup}b), due to the correlations between signal and idler beams.

The two-crystal scheme is equivalent to a nonlinear, or SU(1,1), interferometer~\cite{Chekhova2016}: PDC occurs in the first BBO crystal and is
then amplified or deamplified in the second one, depending on the phase delays in the gap between the two. In particular, at a distance $L$ such that the emission probability amplitudes from the two crystals acquire a relative $\pi$ phase due to the dispersion in the air ($L\approx18$ mm at low pump power), there is full deamplification of the collinear emission~\cite{Perez2014}. The emitted PDC radiation in the far-field zone is then of a typical `donut' shape, as one can see in Fig.~\ref{setup}b.

After the crystals, the pump beam is reflected by a dichroic mirror DM, and the spatial properties of the PDC radiation are analyzed in several ways.


\textit{Measuring far-field intensity distributions (Fig.~\ref{experimental}a).} By placing a CCD camera into the focal plane of the lens L1 (focal distance $f=200$ mm), one can observe intensity distributions as shown in Fig.~\ref{setup}b. The distance $r$ from the center of the distribution is related to the angle of emission $\theta$ as $r=f\theta$ (here and further, we assume $\theta$ to be small). The wavelength spectrum is filtered in this case by placing into the beam an interference filter (IF), with the central wavelength 710 nm and bandwidth 10 nm.

The OAM of the light can be changed by imprinting an azimuthal phase with the help of a spatial light modulator (SLM), which can be placed in the far field of PDC instead of the CCD camera (as shown in Fig.~\ref{experimental}c).

By triggering the CCD camera synchronously with the laser pulses and setting the acquisition time to be less than 1 ms, one obtains single-shot far-field intensity distributions (angular spectra) of PDC. These spectra are further processed for calculating the intensity covariance and for reconstructing the shapes and weights of the spatial modes.

\textit{Measuring the number of spatial modes (Fig.~\ref{experimental}b).} The total number of modes can be assessed through the measurement of the second-order intensity correlation function $g^{(2)}$, rigorously defined through the normally ordered second moment of the photon number, $g^{(2)}=\langle:N^2:\rangle/\langle N\rangle^2$. For large mean photon number, which is the case for BSV, $\langle:N^2:\rangle\approx\langle N^2\rangle$, and the correlation function can be measured by analysing the output current of a single analog detector (D). For a single mode of non-degenerate PDC, $g_0^{(2)}=2$; because in the presence of  $K$ modes the correlation function changes as $g^{(2)}=1+(g_0^{(2)}-1)/K$~\cite{Ivanova2006,Finger2015}, the number of modes can be inferred from the $g^{(2)}$ value:
\begin{equation}\label{g_and_K}
 K=\frac{1}{g^{(2)}-1}.
\end{equation}
 For eliminating the effect of multiple frequency modes, this kind of measurement requires a monochromator selecting fewer than one frequency mode~\cite{Perez2014}. In our experiment, the measurement of $g^{(2)}$ is realized with the monochromator selecting a bandwidth of $0.1$ nm (which is less than the correlation width, see Ref.~\cite{Perez2014}) around the non-degenerate wavelength $708$ nm. The radiation is coupled into the input slit of the monochromator by lens L2 (focal length $25$ mm) and then into the detector by lens L3 (focal length $30$ mm).

Except for the measurement of the total number of spatial modes, all other measurements are performed using a $10$ nm interference filter centered at 710 nm or at $700$ nm if signal-idler correlations have to be eliminated.

\textit{Measuring the weights of OAM modes (Fig.~\ref{experimental}c).} In order to characterize the OAM content of BSV, we use the simplest way of filtering out a single OAM mode, namely the so-called `phase flattening' method~\cite{Mair2001,Qassim2014}. In this method, an SLM is placed into the beam under study and then the diffraction pattern is observed after imprinting an azimuthal phase on the beam. If this procedure compensates for the azimuthal phase of one of the OAM modes, this mode produces a peak in the collinear direction after diffraction. In the original version~\cite{Mair2001}, the beam is then coupled into a fiber, and the coupling efficiency is used to determine the weight of the corresponding OAM mode.
In our experiment, we use a simplified method in which the diffracted beam is observed with a CCD camera. The camera is placed in the focal plane of lens L4, with the focal length $500$ mm. The weight of an OAM mode is inferred from the intensity measured by the CCD camera in the collinear direction (at the center of the intensity distribution), with the averaging made over a square of $3$x$3$ pixels.


\subsection{Calculation of the Schmidt modes}
The concept of the Schmidt modes is crucial for describing photon-number correlations in PDC, FWM, or MI. The probability amplitude of a photon pair emitted into two plane-wave monochromatic modes $\vec{k}_{s,i}$ scales as the so-called two-photon amplitude (TPA) $F(\vec{k}_{s},\vec{k}_{i})$, which can be written as the Schmidt decomposition,
\begin{equation}
F(\vec{k}_{s},\vec{k}_{i})=\sum_n \sqrt{\lambda_n} u_n(\vec{k}_{s})v_n(\vec{k}_{i}).
\label{TPA}
\end{equation}
Here, $\lambda_n$ are the Schmidt eigenvalues, and  $u_n(\vec{k}_{s}),v_n(\vec{k}_{i})$ the Schmidt modes, satisfying normalization and orthogonality conditions. The advantage of introducing them is that, in contrast to plane-wave monochromatic modes, they minimize intermodal correlations: a signal photon emitted into some Schmidt mode $u_n(\vec{k}_{s})$ has its matching idler photon always in a single mode $v_n(\vec{k}_{i})$.
In classical optics, the counterpart of Schmidt modes are coherent modes~\cite{Mandel}, allowing a coherent-mode decomposition of the degree of coherence, very similar to Eq.~(\ref{TPA}).

It has been shown that the Schmidt modes can be also applied to the description of high-gain PDC or FWM where strong twin beams are generated instead of photon pairs~\cite{Sharapova2015}. In particular, the spatial/wavevector Schmidt modes of the `two-crystal' PDC source were considered in detail in Ref.~\cite{Perez2015}. Due to the axial symmetry of the source (Fig.~\ref{setup}), polar coordinates are most convenient for its description. The modes are then functions of the transverse wavevectors $\vec{q}_{s,i}$, given by the lengths $q_{s,i}$ and the azimuthal angles $\phi_{s,i}$, with the index $s,i$ labeling signal and idler beams.

The Schmidt modes have the form~\cite{Sharapova2015,Perez2015}
\begin{equation}
u_{lp}(\vec{q_s})=\frac{\tilde{u}_{lp}(q_s)}{\sqrt{q_s}}e^{il\phi_s},\,\,\,
v_{lp}(\vec{q_i})=\frac{\tilde{v}_{lp}(q_i)}{\sqrt{q_i}}e^{-il\phi_i},
\label{Schmidt}
\end{equation}
where $\tilde{u}_{lp}(q_s),\,\tilde{v}_{lp}(q_i)$ are their radial profiles.
In the degenerate case, as the one of our experiment, the signal and idler radial parts are indistinguishable, and $\tilde{v}_{lp}=\tilde{u}_{lp}$. It is important that each of these modes carries a phase factor $i^{l}$~\cite{Calvo2007}. The weights of the modes are given by the Schmidt eigenvalues $\lambda_{lp}$, normalized as $\sum_{l,p}\lambda_{lp}=1$. The effective number of Schmidt modes is then calculated as $K=[\sum_{l,p}\lambda_{lp}^2]^{-1}$. As the parametric gain of PDC increases, the modes with smaller weights get suppressed while stronger modes survive~\cite{Sharapova2015}, so that the mode weights redistribute to become
\begin{equation}
\Lambda_{lp}=\frac{\sinh^2{\left[G\sqrt{\lambda_{lp}}\right]}}{\sum_{l,p}\sinh^2{\left[G\sqrt{\lambda_{lp}}\right]}}.
\label{redistribute}
\end{equation}
The strength of the parametric interaction is given by the dimensionless parameter $G$, which scales as the quadratic susceptibility, pump peak field, and the total crystal length. To characterize the parametric gain, one measures how the PDC intensity depends on the pump power under narrowband filtering. This procedure yields $G_0=G\sqrt{\lambda_0}$, with $\lambda_0$ being the weight of the strongest Schmidt mode at low gain. In what follows, we will refer to $G_0$ as to the parametric gain.

According to Eq.~(\ref{Schmidt}), all Schmidt modes with $l\ne0$ carry orbital angular momentum~\cite{Allen1992}. Here, we will be interested in all modes (with different $p$ indices) having the same $l$ index. Their number is given by $K_{\rm OAM}=[\sum_{l}\Lambda_{l}^2]^{-1}$, where the weight of an OAM mode is $\Lambda_{l}=\sum_p\Lambda_{lp}$, $\sum_l\Lambda_{l}=1$.

The intensity distribution in the far field is given by the incoherent sum over all Schmidt modes~\cite{Sharapova2015},
\begin{equation}
I(\vec{q})\sim\sum_{l,p}\Lambda_{lp}|u_{lp}(\vec{q})|^2.
\label{intensity}
\end{equation}
As expected, it does not depend on the azimuthal angle $\phi$. The weights of the mode contributions in Eq.~\ref{intensity} are proportional to the mean photon numbers in the modes, $\Lambda_{lp}=\langle N\rangle_{lp}/\langle N\rangle$, where $\langle N\rangle$ is the total mean number of photons. From Eq.~(\ref{intensity}), one can immediately see that if the mean intensity in the collinear direction is zero, which is a consequence of destructive interference for collinear emission, all Schmidt modes have to have `donut' shapes in the far field.

One calculates the Schmidt modes and eigenvalues through the numerical singular-value decomposition of the TPA describing the emission of two photons with the transverse wave vectors $\vec{q}_{s,i}$. This procedure, as well as the derivation of the TPA, is described in detail in Refs.~\cite{Perez2014,Perez2015}. For comparison with the experiment, where one measures the angle of emission $\theta$ instead of the transverse wave vector $q$, we will further use $\theta$ everywhere.The relation between the two values is $\theta=q\lambda/2\pi$, where $\lambda$ is the wavelength of the detected radiation.

\begin{figure}[htbp]
\centering\includegraphics[width=8.5cm]{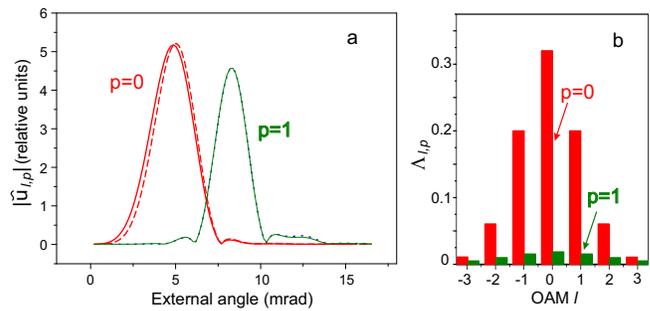}
\caption{Calculated shapes (a) and weights (b) of the strongest radial Schmidt modes for two $2$ mm crystals separated by a distance of $15$ mm. The parametric gain is $G_0=7.6$. Panel (a) shows the absolute values of the modes $\tilde{u}_{0,0}(\theta)$ (red solid line), $\tilde{u}_{3,0}(\theta)$ (dashed red line), $\tilde{u}_{0,1}(\theta)$ (thin solid green line), and $\tilde{u}_{3,1}(\theta)$ (dotted green line). Modes with the same $p$ values and different $l$ values turn out to have very similar shapes; for this reason, we only show the ones for $l=0$ and $l=3$, the profiles of the $l=1,2$ modes being between these two. Panel (b) shows the weights $\Lambda_{l,p}$ of the strongest modes with $p=0$ (red) and $p=1$ (green).}
\label{calc_modes}
\end{figure}

Fig.~\ref{calc_modes} shows the shapes of the few strongest modes calculated for two crystals of $2$ mm length placed at a distance of $15$ mm and pumped at $355$ nm with the pump beam full width at half maximum (FWHM) $170\,\mu$m. The wavelength is assumed to be degenerate, $\lambda=710$ nm, and the value of $G$ in (\ref{redistribute}) is $54$ (which corresponds to the parametric gain $G_0=8.7$). In the calculation, we took into account the Kerr effect for the pump beam (see the next sections for more information).

It should be mentioned that we label the modes in $p$ not according to the number of nodes of the radial profile, as is the case, for instance, for Laguerre-Gaussian (LG) modes, but simply according to the weights: modes with larger $p$ have smaller $\lambda_p$. One can see that the radial shapes of modes with $p=0$ and different $l$ values almost coincide. The same is true for modes with $p=1$ and different $l$. All modes have the shape of 'donuts', in full agreement with the fact that at the chosen distance between the two crystals, contributions of the two crystals into collinear emission interfere destructively. It is noteworthy that the overlap between the different radial modes is very small. This means that one could filter a certain radial mode almost perfectly without introducing considerable losses. This feature is one of the advantages of the chosen `two-crystal' configuration. Indeed, the Schmidt modes of this system are different from the LG modes~\cite{Mair2001} or Hermite-Gaussian modes~\cite{Straupe2011}, which are a good approximation to the eigenmodes of PDC radiation from a single crystal. This is due to the interference structure of the radiation from the `two-crystal' scheme. Clearly, each Schmidt mode as shown in Fig.~\ref{calc_modes} also allows a decomposition in LG modes. However, if addressing certain radial modes is required, this set of modes is more convenient than the LG one where modes with different $p$ have a strong overlap. As to the sorting of the modes in $l$, it can be done using the methods already developed for LG modes.


For each radial mode $p$, azimuthal modes with different $l$ are possible. Their shapes, according to Eq.~(\ref{Schmidt}), are $e^{il\phi}$, as for other types of OAM modes, like LG ones. At the same time, instead of these modes, one can pass to the superpositions of modes with $\pm l$, which have the forms $\sin l\phi$ and $\cos l\phi$. While modes $e^{il\phi}$ with the opposite signs of $l$ should manifest photon-number correlations, their superpositions $\sin l\phi$ and $\cos l\phi$ should manifest quadrature squeezing.

\subsection{Reconstruction of the Schmidt modes}
In a recent paper~\cite{Finger2016}, it was shown that the frequency Schmidt modes of high-gain MI, FWM, and PDC can be reconstructed from the covariance of single-shot frequency spectra. At high gain, this covariance has a structure similar to the one of the squared TPA (\ref{TPA}), except that the coefficients in the Schmidt decomposition are at high gain $\Lambda_n$ rather than $\sqrt{\Lambda_n}$~\cite{footnote1}. Here we apply the same theory to the case of spatial Schmidt modes (\ref{Schmidt}). Further, we will assume that the measured photon numbers have contributions from both signal and idler beams, which is indeed the case in our experiments.

The covariance between the photon numbers emitted into plane-wave modes $\vec{q}$, given by $\theta,\phi$, and $\vec{q'}$, given by $\theta',\phi'$, is defined as
\begin{eqnarray}
\textrm{Cov}(\vec{q},\vec{q'})\equiv\langle [N_s(\vec{q})+N_i(\vec{q})][N_s(\vec{q'})+N_i(\vec{q'})]\rangle-\nonumber\\
-\langle N_s(\vec{q})+N_i(\vec{q})\rangle\langle N_s(\vec{q'})+N_i(\vec{q'})\rangle.
\label{defCov}
\end{eqnarray}
In terms of Schmidt modes $u_{lp}(\vec{q}),v_{lp}(\vec{q})$ from Eq.~(\ref{Schmidt}) it has the form
\begin{eqnarray}
\textrm{Cov}(\vec{q},\vec{q'})=\left[\sum_{l,p} \Lambda_{lp} u_{lp}(\vec{q})u^*_{lp}(\vec{q'})\right]^2+\nonumber\\
+\left[\sum_{l,p} \Lambda_{lp} v_{lp}(\vec{q})v^*_{lp}(\vec{q'})\right]^2+\nonumber\\
+2\left|\sum_{l,p} \Lambda_{lp} u_{lp}(\vec{q})v_{lp}(\vec{q'})\right|^2.
\label{Cov}
\end{eqnarray}

The first two terms in this expression correspond to the auto-correlation of signal and idler radiation and have the form of coherent-mode expansion~\cite{Mandel}, while the third term describes the signal-idler cross-correlation and has the form of the Schmidt decomposition (except that it has the Schmidt eigenvalues instead of their square roots, compare with Eq.~(\ref{TPA}). Usually, one can distinguish between the autocorrelation and cross-correlation parts in the covariance distribution~\cite{Finger2016}. With only auto- or cross-correlation part left, one can retrieve the Schmidt modes through the singular-value decomposition. In the Results section, we consider this procedure separately for radial and azimuthal variables. We will also assume in what follows that $u_{lp}=v_{lp}$, as is the case for the radiation at the degenerate or nearly degenerate wavelength.

\section{Results}
\begin{figure}[htbp]
\centering\includegraphics[width=8.5cm]{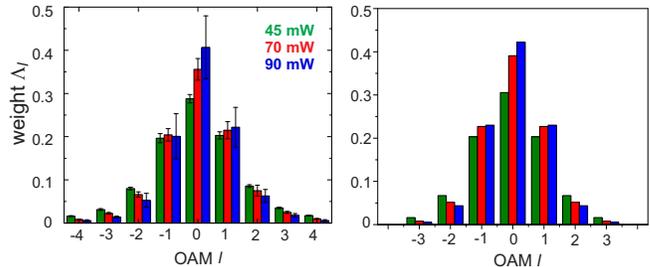}
\caption{The mode weights measured (left) and calculated (right) for different pump power values. The distance between the crystals is 18 mm.}
\label{histogram}
\end{figure}

The weights of the OAM modes $\Lambda_{l}$ are measured using the phase flattening method for three different pump powers, with the distance between the crystals being $18$ mm. The results are presented in Fig.~\ref{histogram}. One can see that, in full agreement with Eq.~(\ref{redistribute}), the mode weight distribution becomes narrower as the pump power increases. The parametric gain depends on the pump power $P$ as $G_0\sim \sqrt{P}$, and we estimate it by measuring the dependence of the output PDC intensity on the pump power. For the distributions presented in Fig.~\ref{histogram}, the gain values found from experiment are $7\pm0.5$, $9\pm0.5$, and $10\pm0.5$.

\begin{figure}[htbp]
\centering\includegraphics[width=7cm]{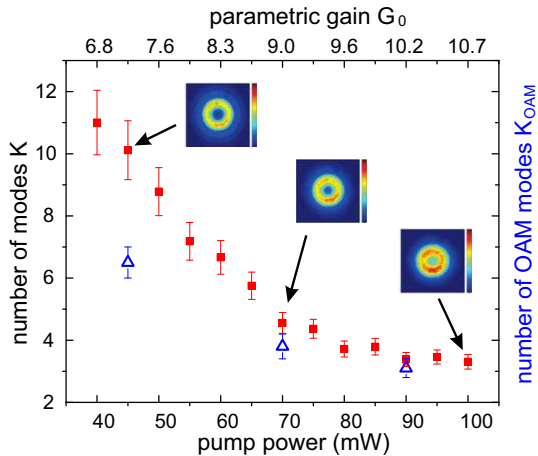}
\caption{The total number of spatial modes $K$ (red) and the number of OAM modes $K_{\rm OAM}$ (blue) measured versus the pump power. The distance between the crystals is $18$ mm.The values of the parametric gain $G_0$ are shown on the top horizontal axis. The insets show the intensity distributions in the far field.}
\label{K(power)}
\end{figure}

The number of OAM modes is found from the distributions in Fig.~\ref{histogram} as $K_{\rm OAM}=1/\sum_l\Lambda_l^2$ and plotted in Fig.~\ref{K(power)} as a function of the pump power. In the same graph, we plotted the total number of spatial modes, determined from the correlation function measurement [see Eq.~(\ref{g_and_K})]. The total number of modes, as expected from Eq.(\ref{redistribute}), is also reduced as the parametric gain increases. Interestingly, while at low gain, the total number of modes exceeds the number of OAM modes approximately twice, at high gain $K\approx K_{\rm OAM}$, which means that for each $l$ there is roughly a single radial mode. In other words, the higher-order radial modes are suppressed stronger with the increase of the gain than the higher-order azimuthal ones. This has a simple explanation: in the chosen geometry of experiment, the number of radial modes at low gain is smaller than the number of azimuthal modes. Therefore it reduces faster as the gain increases, according to Eq.~(\ref{redistribute}).

Note that although the number of spatial modes is measured with narrowband ($0.1$ nm) filtering, leaving only $1.25$ frequency modes, the OAM content is measured with a $10$ nm interference filter. However, because the azimuthal part of the Schmidt modes is nearly independent on the wavelength, this approach is valid.

The insets in Fig.~\ref{K(power)} show the shapes of the intensity distribution in the far-field zone. We see that as the parametric gain increases, the number of observed 'donuts' gets smaller, which is in agreement with the reduction in the number of radial modes. One can also notice that the intensity in the collinear direction increases with the pump power. The reason for this is the Kerr effect: as the pump power increases, the pump beam acquires an additional phase in the crystals, which changes the phase shift in the SU(1,1) interferometer.

\begin{figure}[htbp]
\centering\includegraphics[width=8.5cm]{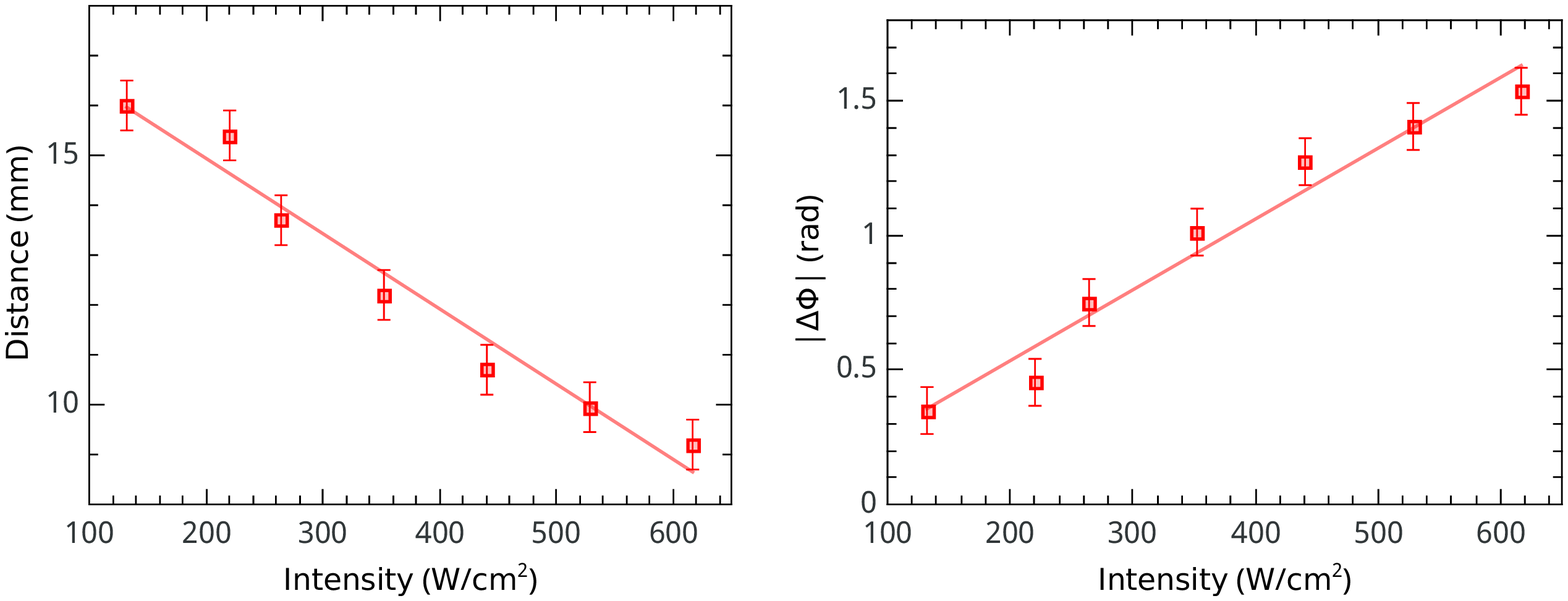}
\caption{The influence of the Kerr effect on the two-crystal scheme. The change in the distance between the two crystals at which complete destructive interference occurs in the collinear direction (left) and the corresponding phase shift induced in the pump beam (right) measured as functions of the pump power. Linear dependences are expected, shown with solid lines.}
\label{Kerr}
\end{figure}

Figure~\ref{Kerr} demonstrates this effect: the left panel shows the pump-dependent change in the distance between the crystals at which we observe a minimum in the collinear emission. At low pump intensity, this distance is $18$ mm - this number is determined by the dispersion of the air~\cite{Perez2014}. As the mean pump intensity increases to $620$ W/cm$^2$ (which means the peak intensity about $100$ GW/cm$^2$), the distance between the crystals corresponding to the destructive interference in the collinear direction is reduced to $9$ mm. Because the output intensity oscillates with a period of $36$ mm regardless of the pump power, from the change in the position of the minimum one can find the additional phase acquired by the pump in the crystal due to the Kerr effect. This additional phase is plotted in the right panel of Fig.~\ref{Kerr} and, as one could expect, it depends linearly on the pump power.

\begin{figure}[htbp]
\centering\includegraphics[width=6cm]{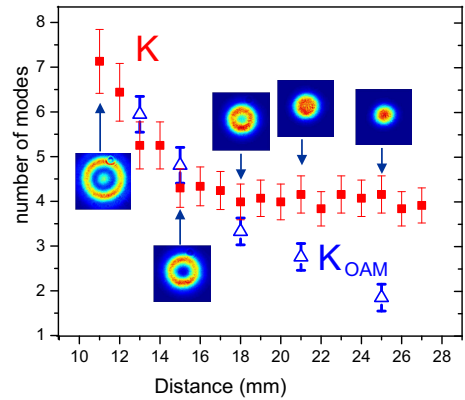}
\caption{The total number of spatial modes (red squares) and the number of OAM modes (blue triangles) versus the distance between the crystals for the pump power $90$ mW. The insets show the intensity distributions.}
\label{modes(L)}
\end{figure}

As the distance between the crystals is varied, the total number of spatial modes changes~\cite{Perez2014}. This tendency is shown as red points in Fig.~\ref{modes(L)} where the intensity distributions are also shown as insets. This measurement demonstrates a monotonic decrease of both the total number of modes and the number of OAM modes (blue triangles) as the distance increases from $12$ to $25$ mm.

At $L<18$ mm, the total number of modes approximately coincides with the number of OAM modes, which means that only modes with a single value of $p$ contribute to the spectrum, in agreement with the theoretical calculation (Fig.~\ref{calc_modes}). On the other hand, at distances above $18$ mm the number of spatial modes is approximately constant, while the number of OAM modes is reduced. The reason is that with the increase of the distance between the crystals, destructive interference in the collinear direction gradually turns into constructive interference, and a strong intensity peak appears at the center. This peak is only compatible with modes having no OAM, $l=0$. Indeed, any mode with $l\ne0$ has a phase singularity connected with the OAM content and therefore does not allow for intensities along the beam axis. On the other hand, emission at large angles is not amplified in the second crystal because it has a poor overlap with the pump beam~\cite{Perez2014}. For both these reasons, OAM modes are naturally suppressed in this region.

Finally, we present the reconstruction of the mode shapes from the single-shot angular spectra) as shown in Fig.~\ref{setup}b. These spectra are given by
\begin{equation}
I(\theta,\phi)\sim\sum_{l,p}N_{lp}|u_{lp}(\theta,\phi)|^2,
\label{single_pulse}
\end{equation}
which differs from (\ref{intensity}) only in that it contains random numbers $N_{lp}$ of photons populating each Schmidt mode rather than mean numbers. In other words, different modes contribute to these distributions with random weights. Note that the averaging of (\ref{single_pulse}) over time (different pulses) yields (\ref{intensity}).

In the experiment, $3500$ single-shot two-dimensional (2D) angular spectra are recorded with a CCD camera.
 To eliminate the effect of the pump intensity fluctuations, each spectrum is normalized to the integral intensity.

First, for reconstructing the radial modes we fix the azimuthal angle $\phi$ within a certain narrow range around $0$. Namely, in each spectrum $S(\theta,\phi)$ integration is performed over a range of $\phi$ from $-4.5^\circ$ to $4.5^\circ$~\cite{footnote2}. For the resulting one-dimensional spectra $S(\theta)$, the covariance is calculated as
\begin{equation}
\textrm{Cov}_S(\theta,\theta')=\langle S(\theta)S(\theta')\rangle-\langle S(\theta)\rangle\langle S(\theta')\rangle.
\label{Cov_exp}
\end{equation}
In order to include the cross-correlation part of the covariance, which peaks at the signal and idler azimuthal angles differing by $\pi$, we have also introduced negative values of $\theta,\theta'$, corresponding to the values of $\phi$ around $\pi$.

\begin{figure}[htbp]
\centering\includegraphics[width=8.5cm]{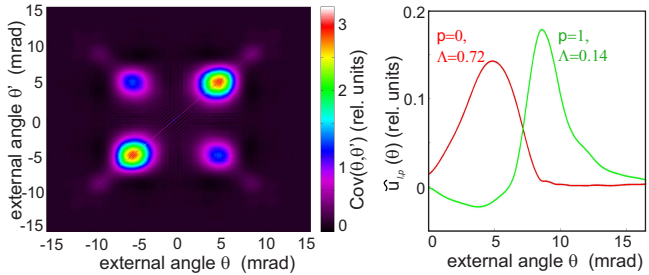}
\caption{The covariance (left) measured from single-shot radial spectra and the two strongest radial modes (right) reconstructed from it. The distance between the crystals is $15$ mm, the power is $45$ mW. Single-shot spectra are integrated over $9^\circ$ of azimuthal angle.}
\label{radial}
\end{figure}

The covariance obtained in this manner is shown in the left panel of Fig.~\ref{radial}.
Clearly, it has two contributions: the auto-correlation one, localized in the top right and bottom left quadrants, and the cross-correlation one, localized in the other two quadrants. The auto-correlation peaks are more pronounced than the cross-correlation ones, probably because the filter central wavelength $710$ nm is shifted from the degenerate wavelength $709.3$ nm and the uncompensated frequency modes only contribute to auto-correlation.

Both the auto-correlation part and the cross-correlation part of the covariance can be used for extracting the radial Schmidt modes. Here, we used the auto-correlation part, corresponding to $\phi=\phi'$ in Eq.~(\ref{Cov}), and took into account the frequency degeneracy as well as the fact that the radial mode shapes $\tilde{u}_{lp}$ do not depend on $l$ (see Fig.~\ref{calc_modes}a).~\footnote{This statement is valid only for the chosen geometry and for $p\le3$; however, modes with larger $p$ have weights less than $0.001$.} Then, the auto-correlation part takes the form
\begin{eqnarray}
\textrm{Cov}_S(\theta,\theta')|_{\phi=\phi'=0}\sim\left[\sum_{p=0}^{\infty} \Lambda_{p} \tilde{u}_{p}(\theta)\tilde{u}^*_{p}(\theta')\right]^2,\nonumber\\
\Lambda_{p}\equiv\sum_{l=-\infty}^{\infty}\Lambda_{lp}.
\label{Cov_theta}
\end{eqnarray}

From the square root of this distribution, the shapes and weights of the radial modes are found through the singular-value decomposition. The shapes and weights of the first two radial modes are shown in the right panel of Fig.~\ref{radial}. The weights of the $p=0$ and $p=1$ modes are found to be, respectively, $0.72\pm0.05$ and $0.14\pm0.05$. This slightly differs from the calculated data (Fig.~\ref{calc_modes}, where after summing over the weights of modes with different $l$, one obtains $0.86$ for $p=0$ and $0.09$ for $p=1$). The shapes of the modes are in a good agreement with the calculated ones.


For reconstructing the azimuthal (OAM) mode structure, one has to get rid of the radial functions in Eq.~(\ref{Cov}). It is worth mentioning that the azimuthal Schmidt modes have the trivial shapes $e^{il\phi_{s,i}}$ simply due to the periodicity of the TPA in $\phi_s-\phi_i$. Therefore, the only parameter of interest is the distribution of eigenvalues $\{\Lambda_l\}$. For finding it, we first use the property of the radial modes for the two-crystal configuration: namely, that the radial functions of different orders overlap very little, $u_{p}(\theta)u_{p'}(\theta)\sim\delta_{pp'}$. This theoretical result is confirmed by the experimental data (Fig.~\ref{radial}). Then, by picking an angle $\theta_0$ corresponding to a mode with a certain $p=p_0$, we obtain from Eq.~(\ref{Cov})
\begin{eqnarray}
\textrm{Cov}(\phi,\phi')=2\left[\sum_{l=-\infty}^{\infty}\Lambda_{lp_0} |u_{p_0}(\theta_0)|^2e^{il(\phi-\phi')}\right]^2+\nonumber\\
+2\left|\sum_{l=-\infty}^{\infty}\Lambda_{lp_0} |u_{p_0}(\theta_0)|^2e^{il(\phi-\phi'+\pi)}\right|^2,
\label{Cov_phi}
\end{eqnarray}
the first and the second terms corresponding to auto- and cross-correlation, respectively.

In the experiment, for each single-shot spectrum we integrate over the angle $\theta$ within the limits of $\pm1.1$ mrad around the maximum of the main `donut', $p_0=0$. The covariance of the resulting spectra $S(\phi,\phi')$ depends on $\Delta\phi\equiv\phi-\phi'$ as
\begin{equation}
\textrm{Cov}_S(\phi,\phi')\sim\left[\sum_{l=-\infty}^{\infty}\Lambda_{l0} e^{il\Delta\phi}\right]^2+\left[\sum_{l=-\infty}^{\infty}\Lambda_{l0} e^{il(\Delta\phi+\pi)}\right]^2,
\label{Cov_phi_exp}
\end{equation}
where the modulus in the last term is omitted because this term is real.

\begin{figure}[htbp]
\centering\includegraphics[width=8.5cm]{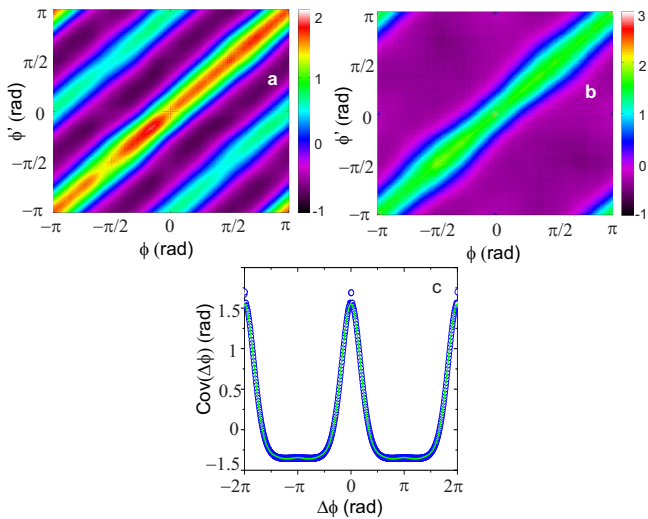}
\caption{Panel a: the covariance calculated from single-shot spectra. The distance between the crystals is $15$ mm, the power is $45$ mW. Single-shot spectra are integrated over $2.2$ mrad of the polar angle $\theta$. Panel b: the same covariance, calculated from the single-shot spectra obtained with the interference filter centered at $700$ nm, which leads to the disappearance of the `cross-correlation' stripes. Panel c: the covariance from panel b as a function of $\Delta\phi=\phi'-\phi$, averaged over $\phi+\phi'$ (blue points). Green line shows the fit with the first term of (\ref{Cov_phi_exp}).}
\label{azimuthal}
\end{figure}

This distribution is shown in Fig.~\ref{azimuthal}a. The central stripe corresponds to the auto-correlation (the first term in Eq.~(\ref{Cov_phi_exp})) and the two other stripes, to the cross-correlation (the second term). Note the qualitative resemblance of the covariance plot with the TPA calculated in Ref.~\cite{Fedorov2016}, although for a different (noncollinear) geometry. One can also notice that, contrary to the structure of Eq.~(\ref{Cov_phi_exp}), the distribution has negative values. This can be explained by the fact that, by normalizing each single-shot frame to its integral, we force the total intensity to be constant. This is incompatible with the intensity fluctuations due to the relatively small number of radial and frequency modes and causes anti-correlations, i.e., covariance negativities.

In order to extract the OAM mode weights $\Lambda_{l0}$ from the measured covariance, it is convenient to get rid of one of the terms in Eq.~(\ref{Cov_phi_exp}). Since the autocorrelation term cannot be avoided, we eliminate the cross-correlation term by recording experimental spectra as shown in Fig.~\ref{experimental}a with a $10$ nm interference filter centered at $700$ nm. This way we get rid of the signal-idler correlations, and in the resulting covariance (Fig.~\ref{azimuthal}b), only the auto-correlation `stripe' is present. Further, because the covariance should not depend on the sum of $\phi,\phi'$ but only on their difference, we perform averaging of the distribution in Fig.~\ref{azimuthal} along the $\phi+\phi'$  direction. The resulting distribution is shown in Fig.~\ref{azimuthal}c by the blue points. We obtain the $\{\Lambda_{l0}\}$ values by fitting this dependence with the first term of Eq.~(\ref{Cov_phi_exp}) where a negative background is introduced. The green line shows the fit, with the weights $\Lambda_{l0}=\Lambda_{-l0},\, l=0,1,2,3$ as fitting parameters. The obtained values $0.28,0.21,0.05,0.03$ agree well with the theoretically calculated ones (Fig.~\ref{calc_modes}b).

\section{Discussion}
The two-crystal configuration used in this work turns out to have some important advantages. Because of the destructive interference in the collinear direction, all Schmidt modes have `donut' shapes, with donuts of different radiuses having different radial numbers. The strongest of them are almost non-overlapping, which enables their easy sorting. At the same time, for each radial `donut' mode, there is still a rich spectrum of OAM modes. In particular, the strongest mode has $l=0$, which at first sight seems counter-intuitive. Indeed, in `standard' sets like LG ones, the mode with $l=0$ is Gaussian, while any mode with nonzero OAM has zero intensity at the center. But the inverse is not true: a donut-shaped beam can have no OAM.

Note that the modes described here are not invariant under the Fourier transformation and therefore the near-field modes have shapes different from the far-field ones. Their investigation, however, is outside of the framework of this paper.

We see that the effective (Schmidt) numbers of both spatial modes in general and the OAM modes in particular decrease with the gain, due to the usual competition of Schmidt modes, reported in earlier works. At the same time, this decrease is stronger for the radial modes than for the OAM modes. The reason is that even at low gain, in the chosen geometry the number of radial modes is smaller than the number of azimuthal (OAM) modes. This asymmetry becomes even more pronounced at high gain, and the number of radial modes reduces faster than the number of OAM modes.

The numbers of both the radial modes and the OAM modes can be modified by varying the distance between the two crystals. It is noteworthy that the OAM Schmidt number is reduced drastically as the interference becomes constructive for the collinear direction. The largest effective number of OAM modes achieved so far is 6, about the same as the total number of spatial modes, and corresponds to the distance between the crystals $13$ mm.

To make the OAM spectrum richer, one could put a lens between the two crystals and `image' the radiation from the first crystal on the second one. In this case, one can expect no reduction in the total angular width of the spectrum, but only the modulation of this spectrum. Similarly, at a certain distance between the crystals the collinear emission will be suppressed, which will simplify the OAM manipulation. In addition, higher OAM values can be achieved.

Instead of focusing on the OAM modes $e^{il\phi}$, one can consider their superpositions, $\cos(l\phi)$ and $\sin(l\phi)$. These modes are also known as `petal beams' and can be used for imaging and communications~\cite{fickler2013real}. As the OAM modes with opposite $l$ values should manifest photon-number correlations, their sine and cosine superpositions should manifest quadrature squeezing. However, their study, as well as the study of photon-number correlations in OAM modes, is outside of the scope of this paper.

\section{Conclusion}
We have considered the two-crystal scheme from the viewpoint of generating bright squeezed vacuum in OAM modes. The scheme has turned out to be a versatile source of OAM modes populated by large numbers of photons. It provides non-overlapping radial Schmidt modes and therefore one can expect that it enables easy sorting of the modes.

We demonstrated variation of the number of OAM modes from 2 to 6 by changing the distance between the crystals and from 3 to 6 by changing the pump power. More flexibility is expected with using advanced geometries, such as a lens placed between the two crystals and imaging the near field of the first one on the second one.

Finally, to the best of our knowledge, for the first time the reconstruction of the spatial Schmidt modes shapes and weights from the covariance measurement on single-shot spectra has been reported. This method can be successfully applied in classical optics, for instance, for the reconstruction of the spatial modes of optical fibers.

\section{Funding}
This work was funded by the Deutsche Forschungsgemeinschaft (DFG) - CH1591/2-1. R.F. acknowledges the support of the Banting postdoctoral fellowship of the Natural Sciences and Engineering Research Council of Canada (NSERC). R.F. and R.W.B. gratefully acknowledge the support of the Canada Excellence Research Chairs Program (CERC), the Natural Sciences and Engineering Council of Canada (NSERC), and the University of Ottawa Max Planck Centre for Extreme and Quantum Photonics.


\begin{thebibliography}{99}
\bibitem{footnote2}{A smaller integration interval leads to more noise but does not change the result considerably.}
\end{thebibliography}
\end{document}